\documentclass[10pt]{article}
\usepackage{opex3_arxiv}
\usepackage{varioref}
\usepackage{graphicx}
\usepackage[T1]{fontenc}
\usepackage{epsfig} 
\usepackage{subfigure}
\usepackage{cite}
\usepackage{amsmath}
\usepackage{amsfonts}
\usepackage{stmaryrd}
\usepackage{psfrag}
\usepackage{tabularx}\newcolumntype{Y}{>{\centering\arraybackslash}X}
\usepackage[french]{babel}
\usepackage{epstopdf}
\graphicspath{{/home/image/eps/pmma/}}
\DeclareGraphicsExtensions{.eps,.EPS,.ps,.PS,.eps.gz}
\begin{document}

This paper has been published in Opt. Express in 2013 under the reference :\\

T. Grosjean, M. Mivelle, G. Burr, and F. Baida, Opt. Express  21, 1762-1772 (2013). \\

\newpage

\title{Fiber-integrated single photon source of high efficiency based on a concept of ultra-broadband optical horn antenna}

\author{T. Grosjean,$^{1*}$ M. Mivelle,$^{1,2}$ G.W. Burr$^2$ and F.I. Baida$^1$}

\address{$^{1}$ Département d'Optique P.M. Duffieux,\\ Institut
FEMTO-ST, UMR CNRS 6174, Universit\'e  de Franche-Comt\'e,\\ 16
route de Gray, 25030 Besançon cedex, France}

\address{$^{2}$ IBM Almaden Research Center, D2/K13E,\\ 650 Harry
Road, San Jose, California 95120, United States of America}

\email{*thierry.grosjean@univ-fcomte.fr}

\begin{abstract}
We theoretically demonstrate a fiber-integrated single photon source of unprecedented efficiency. This fiber single photon source is achieved  
by coupling optically a single quantum emitter to a monomode optical fiber with a new concept of ultra-broadband optical antenna. Such an optical antenna concept is the result of the transposition to optical frequencies of the well-known low-frequency horn antenna  The optical horn antenna is here shown to be capable of directing the radiation from the emitter toward the optical fiber and efficiently phase-matching the photon emission with the fiber mode. Numerical results show that an optical horn antenna can funnel up to 85\% of the radiation from a dipolar source within an emission cone semi-angle as small as 7 degrees (antenna directivity of 300).  It is also shown that 50\% of the emitted power from the dipolar source can be collected and coupled to an SMF-28 fiber mode over spectral ranges larger than 1000 nm, with a maximum energy transfer reaching 70 \%. This approach is highly promising in the engineering of all-fiber "`on-demand"' single photon sources for telecommunication, quantum optics and sensing.
\end{abstract}



\section{Introduction}

Collecting and propagating photons from single quantum emitters (QE) over long distances and complex optical networks is a key issue in quantum optics and metrology, and one that is still highly challenging. Optical fibers often appear to be the best solution to transmit photons with minimum losses. The major problem which hinders the development of in-fiber propagation of the fluorescence signal from nanoscale single emitters is the low coupling efficiency between the omni-directional dipolar emission of the emitter and the low numerical aperture (NA) of optical fibers. Only a small part of the radiated photons can be collected and guided into the fiber. For example, propagation of single photons from isolated quantum dots has been achieved at telecom wavelengths with a source-fiber optical coupling lower than one percent \cite{miyazawa:jjap05}. Various micro and nano-devices \cite{moreau:apl01,pelton:prl02,claudon:natphot10,gerard:josab09,devilez:nl10,curto:science10,lee:natphot11,chen:ol11} have been proven to notably enhance the emission directivity of single fluorophores embedded in or attached to the structures. However, at least in these preliminary studies, efficient light collection is achieved in free space through high-NA objectives and the proposed devices do not seem to be well-adapted to efficiently launch and phase-match photons into low-NA optical fibers.

Recently, the evanescent coupling between optical fibers and metallic wires  \cite{chang:prl06} or semiconductor waveguides \cite{davanco:ol09} has been proposed as an optical communication way to efficiently transfer energy from single QEs to optical fibers. Collection efficiencies as large as 70\% into both opposite propagation channels of a monomode optical fiber have been predicted with this technique (35\% heads upstream, and 35\% downstream)\cite{davanco:ol09}. In this paper, we introduce and theoretically investigate an alternative approach, based on the transposition of the concept of the microwave horn antenna (HA) to optical frequencies \cite{balanis}. It is shown that the emission diagram of the optical emitter coupled to the HA is highly directed toward the core of an optical fiber and the photons launched within the fiber core are efficiently phase-matched with the optical mode, leading to maximum collection efficiencies larger than 70 \% into a single guided propagation channel of a monomode optical fiber. Spectral broadband operation is also predicted with QE-to-fiber coupling efficiencies above 50 \% over spectral bandwidths exceeding 1000 nm in the near-infrared domain. Therefore, optical horn antenna  may relax the spectral matching requirements with narrow emission line nano-emitters and can be used across a large range of emission wavelengths, making the optical HA of high practical interest in many applications such as telecommunications.

\section{Principle and optical antenna geometry}

Horn antennas (HA) form a well-known family of versatile directional electromagnetic emitters and receivers at microwave frequencies \cite{balanis}. The HA transforms the impedance of a metallic hollow waveguide to the impedance of free space (377 $\Omega$) simply by flaring the waveguide into a larger opening. Usually, the HA is fed with a coaxial cable thanks to a coax-to-waveguide transition and the overall structure (flared waveguide+coax-to-waveguide transition) is also often called HA. This work will focus onto this HA configuration, sketched in Fig. 1(a). It consists of a piece of monomode metallic hollow waveguide, closed at one end by a reflector (metal), flaring into an open-ended conical or pyramidal shaped horn at the other end. The structure is fed with the central conductor of the coaxial cable which protrudes within the waveguide. This short piece of thin protruding wire is placed in between the mirror and the tapered waveguide and radiates within the waveguide as a subwavelength (dipolar) microwave source. The distance between the mirror and the feed is generally close to $\lambda_g/4$ where $\lambda_g$ is the effective wavelength of the waveguide mode. This optimum distance is defined precisely from a complex analysis of impedance matching between the coax and the waveguide and is found empirically as a function of the length of source wire \cite{wade:06}. A general rule is that a $\lambda/4$ protruding source wire placed within a waveguide of square section induces a maximum radiation in free space through the HA when it is set at $\lambda_g/4$ from the reflector. The basic principle of the HA is that the back-reflected radiation along the closed channel interferes constructively with the radiation that propagates directly toward the open channel, leading to an overall radiation pattern directed outside of the waveguide. Then, the tapered horn provides a gradual transition structure to match the impedance of the waveguide to the vacuum. Note that high directivity and impedance-matching are achieved for tapered waveguides of aspect ratios (length/aperture size) that are high enough to induce output waves with nearly constant phase across the aperture (planar wavefronts).
\begin{figure}[htbp]
\begin{center}
\includegraphics [width=0.95\columnwidth]{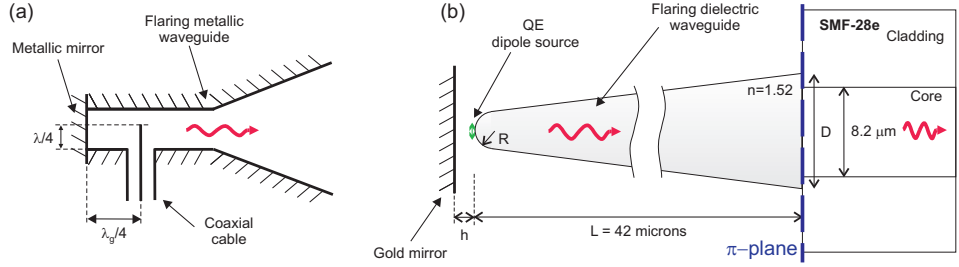}
\caption{(a) Scheme of the microwave horn antenna (b) Scheme of the fiber-integrated optical horn antenna whose architecture follows the one of the microwave HA: a subwavelength dipole source is placed in between a gold mirror and a flaring dielectric waveguide (n=1.52). The dipolar source is set in contact to the dielectric waveguide of length $L=42$ microns. The spacing between the source and the mirror ($h$) is close to $\lambda/4$ for optimum performances of the optical antenna.}\label{fig:scheme}
\end{center}
\end{figure}

We show here that the transposition to optics of this antenna concept can bring an innovative solution to achieve directive emissions from nanoscale light sources which efficiently match the impedance of single mode fibers.  The proposed antenna configuration, depicted in Fig. \ref{fig:scheme}(b), consists of the combination of a tapered dielectric microtip and a gold mirror. These two elements correspond to the flaring waveguide and the reflector of the HA shown in Fig. 1(a), respectively. A point-like optical source (fluorescent molecule, colloidal quantum dot, etc) with transition dipole moment oriented perpendicularly to the axis of symmetry of the HA (parallel to the mirror interface) is placed in between the mirror and the tapered dielectric microtip, in contact of the rounded apex of the microtip. The nanoscale emitter is not embedded in a dielectric nanowire as it could be imagined in a direct transposition to optics of the HA concept. We rather exploit the ability of dielectric interfaces to modify the dipolar radiation pattern of QEs \cite{lukosz:josa77,gerard:josab09} to couple photons from the emitter to the tapered waveguide.  Note that the use of metallic hollow waveguides does not seem to be technologically realistic at optical frequencies and the lossy nature of metals at this spectral range may induce dissipation of the optical energy and therefore limit the quantum efficiency. This configuration of HA can be placed in contact to the cleaved entrance facet of an optical fiber (called $\pi$-plane in the following).

\section{Simulation}

We cascade two different numerical approaches to model the radiation and fiber-coupling processes of our HA prototype. The first HA collection step (light scattering by the QE inserted in the antenna and photon channeling within the flaring dielectric waveguide) is numerically investigated using the BOR-FDTD method (Body-of-Revolution Finite Difference Time Domain)\cite{taflove:book}. The QE is here approximated by a two-level system whose emission transition is governed by a pure electric dipole moment. Its radiation properties can be obtained using the semiclassical approach in the weak-coupling regime by treating the emitter as a classical dipole oscillating at the transition frequency $\omega$ \cite{novotny:nanooptics}. In the following, the QE is assumed to be a perfect emitter with an intrinsic quantum yield of 1. It is defined numerically by a fixed current density at a single cell of the simulation meshgrid. The BOR-FDTD method is limited to the modeling of rotationally symmetric structures with point-like sources positioned along their symmetry axis. The optical fields are then calculated in cylindrical coordinates ($r,\theta,z$) under the form $\textbf{f}(r,z) \exp[i m\theta]$, where $m \in N$. Constant $m$ is the rotational symmetry order of the simulation defined by the electromagnetic spatial symmetry of the source. In our case, the excitation of the HA with an oscillating dipole oriented along the plane ($r,\theta$) perpendicular to the symmetry axis (0$z$) of the structure imposes $m=1$, leading to the simulation of linearly-polarized transverse fields.
The flaring dielectric waveguide on which the QE is attached is placed at a distance $h$ from the mirror (Fig. \ref{fig:scheme}). Its refractive index ($n$) and length ($L$) are chosen to be equal to 1.52 and 42 microns, respectively. Its radius of curvature ($R$) is chosen to be a few hundred nanometers large to efficiently outcouple photons from the dipole source into its tapered body. Dispersion within the dielectric horn is neglected whereas the dielectric constant of the metallic mirror is defined by a Drude model that fits the dielectric constant of gold at $\lambda=1.55$ microns given by Palik's book \cite{palik:book}. Since the computation volume terminates at the end facet of the dielectric horn, the fields calculated at its upper limit can be directly used for the calculation of the mode coupling into a fiber. The second step, of the HA-to-fiber optical coupling, is simulated with the well-known overlap integral method \cite{thomson:83}. Note that the reflection at the $\pi$-plane is not included in our antenna model. Since the index difference between the dielectric horn and the optical fiber does not exceed 0.053, the reflectance at the $\pi$-plane (of the order of 4 10$^{-4}$) is small enough to be neglected.

The photon transfer from the QE to the optical fiber is described by coefficient $T$ defined as :
\begin{center}
\begin{equation}
T=q \eta C_m,\label{eq:T}
\end{equation}
\end{center}

where $q$ is the quantum yield of the emitter, $\eta$ is the collection efficiency of the HA and $C_m$ is the coupling efficiency between the photons that leave the HA and the fiber mode. Factor $T$ represents the ratio of emitted photons that are collected by the HA and outcoupled into the fiber mode. $q$ is obtained from :
\begin{center}
\begin{equation}
q=\frac{\gamma_{r}}{\gamma_{r}+\gamma_{nr}},
\end{equation}
\end{center}

where $\gamma_{r}$ and $\gamma_{nr}$ are the radiative and non radiative decay rates of the QE. We have $\gamma_{r}=P_{r}/P_0$ and $\gamma_{nr}=P_{nr}/P_0$ where $P_{r}$ is the power radiated in free space by the emitter coupled to the HA, $P_{nr}$ is the power dissipated by the HA, and $P_0$ is the power radiated by the emitter in free space without the presence of the antenna.

The collection efficiency ($\eta$) of the HA is defined as the fraction of radiated power that is channeled within the body of the flaring dielectric waveguide. It is obtained from:
\begin{center}
\begin{equation}
\eta=\frac{P_i}{P_r} \label{eq:eta}
\end{equation}
\end{center}
where $P_i$ is the power collected by the HA.

The energy transfer between the collected photons that leave the HA and the fiber mode can be described in terms of the coupling coefficient $C_m$, defined as:
\begin{center}
\begin{equation}
C_m = P_m / P_i
\end{equation}
\end{center}
where $P_m$ and $P_i$ are the powers carried by the fiber mode and the collected optical field that leaves the HA, respectively. $P_m$ and $P_i$ are obtained from Poynting vector integrations over specific areas in the $\pi$-plane:
\begin{eqnarray}
P_i & = & \frac{1}{2} \mathfrak{Re} \iint \, r \, dr \, d\theta \, (\vec
E_i \times \vec H_i^{\ast}) \cdot \vec e_z, \label{eq:pi}\\ P_m & = &
\frac{1}{2} \mathfrak{Re} \iint \, r \, dr \, d\theta \,  (a_m \vec E_m
\times b_m^{\ast} \vec H_m^{\ast}) \cdot \vec e_z. \label{eq:pm}
\end{eqnarray}
Here the unit vector $\vec e_z$ defines the direction of the fiber axis, and the expressions of constants $a_m$ and $b_m$ are based on the overlap integrals between the collected fields that leave the HA ($\vec E_i, \vec H_i$) and the mode field distribution ($\vec E_m, \vec H_m$):
\begin{eqnarray}
a_m & = & \frac{\iint \, r \, dr \, d\theta \,  (\vec E_i \times \vec
H_m^{\ast}). \vec e_z}{\iint \, r \, dr \, d\theta \,  (\vec E_m \times \vec
H_m^{\ast}). \vec e_z}, \label{eq:am}\\ b_m & = & \frac{\iint \, r \, dr \, d\theta \,
(\vec E_m^{\ast} \times \vec H_i). \vec e_z}{\iint \, r \, dr \, d\theta \,  (\vec E_m^{\ast} \times \vec H_m). \vec e_z}.
\label{eq:bm}
\end{eqnarray}
In these simulations, a SMF-28 optical fiber is considered (single mode at telecommunication wavelengths) whose core diameter is 8.2 $\mu$m. Its core and cladding indices are deduced from the mode optical parameters given in the fiber datasheet. For accurate simulation of the HA optical properties, the lateral extent of the FDTD computation volume was chosen to be wide enough (15 microns along the radial coordinate) that the captured field distributions were significantly wider than the lateral extent of the fiber core.

Another critical parameter that describes the performances of optical HAs (and antennas in general) is the directivity, defined as \cite{balanis}:
\begin{center}
\begin{equation}
Dir = max \left( \frac{4 \pi P(\beta,\theta)}{P_{r}} \right), \label{eq:dir}
\end{equation}
\end{center}

where $P(\beta,\theta)$ is the angular power radiated in the direction of polar angle $\beta$ and azimuthal angle $\theta$, and $P_{r}=\iint P(\beta,\theta) sin \beta d\beta d\theta$ is the integral over all angles. An isotropic source would have a directivity of 1, whereas for a dipolar emitter Dir=1.5.

\section{Results and discussions}

\begin{figure}[htbp]
\begin{center}
\includegraphics [width=0.98\columnwidth]{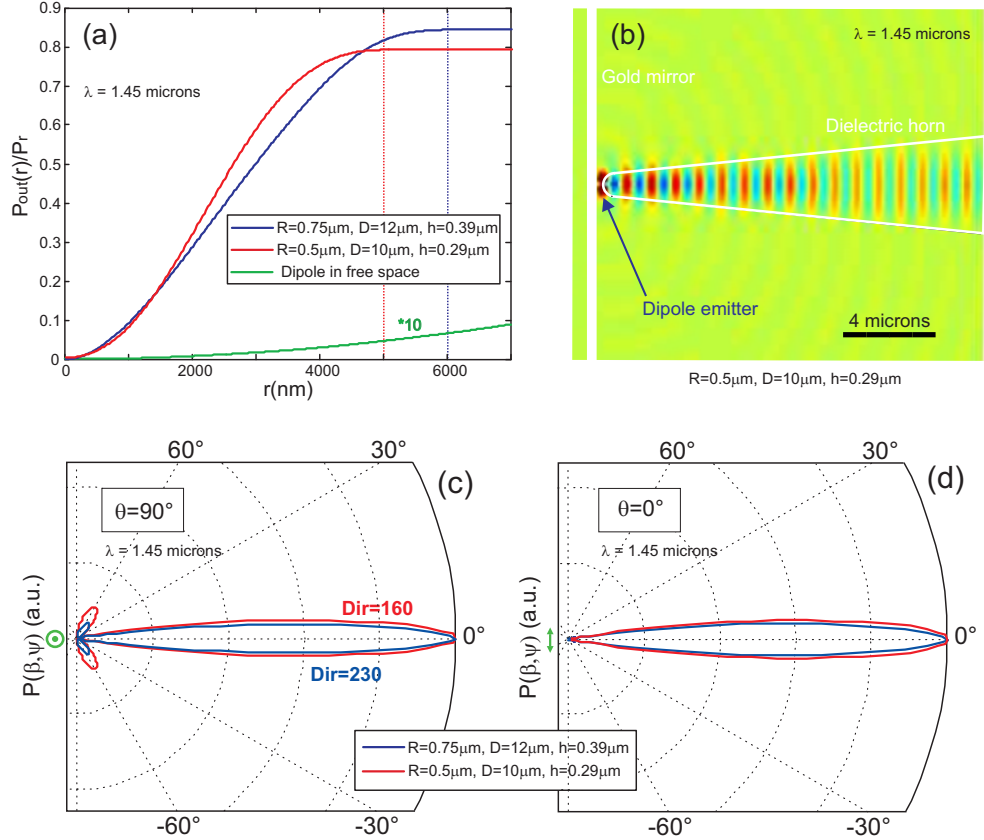}
\caption{(a) Power accumulated by two different geometries of optical HAs (described in the inset) through the output $\pi$-plane, as a function of the radial coordinate $r$. For each HA geometry, the outer boundary of the dielectric horn is represented by a dotted line. (b) Spatial distribution of the real part of the electric field emitted by a dipole source coupled to an optical HA whose geometry is detailed under the figure. The field is displayed within the longitudinal plane ($r,z$) that contains the dipole and the first 16 microns of the flaring dielectric waveguide. (c,d) emission diagrams $P(\beta,\theta)$ when (c) $\theta=90^{\circ}$ (($y,z$)-plane perpendicular to the dipole direction) and (d) $\theta=0^{\circ}$ (($x,z$)-plane that contains the dipole direction), for the two different HA geometries considered in Fig. \ref{fig:colldir}(a) (detailed in the inset). The dipole is shown in green color.}\label{fig:colldir}
\end{center}
\end{figure}

First, let us focus onto the collection properties and directivity of the optical HA. The blue and red curves of Fig. \ref{fig:colldir}(a) show the fraction of radiated power $P_{out}(r)/P_r$ that crosses the output $\pi$-plane of two different HA geometries, as a function of the bounding radial space coordinate $r$. $P_{out}(r)$ is the power transmitted inside the circular area of radius $r$ contained in the $\pi$-plane. The origin of coordinate $r$ is thus the intersection point between the $\pi$-plane and the symmetry axis of the optical HA. The two HA geometries are given in the figure inset and the wavelength $\lambda$ is equal to $1450$ nm.  We see that the two HAs collect 80\% and 85\% of the total power radiated by the dipole source, respectively. The two vertical dotted lines represent the lateral limit in the $\pi$-plane of the two dielectric horns considered here (lines positioned at $D/2$). As a comparison, the green curve of Fig. \ref{fig:colldir}(a) represents $P_{out}(r)/P_r$ achieved in the $\pi$-plane without the presence of HA (free space dipole emission). The collection and emission properties of the optical HA are also revealed in Fig. \ref{fig:colldir}(b) which plots the spatial distribution of the real part of the electric field that propagates along the longitudinal plane ($r,z$) that contains the dipole source and the first 16 microns of the dielectric horn. The real part of the electric field is represented to show the amplitude and phase properties of the waves that are channeled within the dielectric horn. The configuration of optical HA under consideration here is detailed under the figure. We see that the propagating optical field within the flaring dielectric waveguide show planar wavefronts and its amplitude distribution is of gaussian-like shape. This is confirmed by Fig. \ref{fig:colldir}(a) which show accumulated powers in the $\pi$-plane as gaussian-like functions of the radial coordinate. Such field properties may insure highly directive propagations in free space or in optical fibers.

Figures \ref{fig:colldir}(c,d) display the free-space emission diagrams $P(\beta,\theta)$ of the two HA geometries of interest (detailed in the figure insets). The emission diagrams are plotted at $\lambda=1450$ nm for two different values of the azimuthal angle: $\theta=90^{\circ}$ (Fig. \ref{fig:colldir}(c)) and $\theta=0^{\circ}$ (Fig. \ref{fig:colldir}(d)), defining the two perpendicular longitudinal planes ($y,z$) and ($x,z$), respectively. The dipole, represented in green in the figures, is oriented along ($0x$). In these simulations, the optical HA is not considered to be in contact to the optical fiber but to emit directly in vacuum. Figures \ref{fig:colldir}(c,d) show that the photons collected by the two HA configurations are radiated within cone semi-angles of 12$^{\circ}$ (red curves) and 13.5$^{\circ}$ (blue curves), with antenna directivities of 160 and 230, respectively (Eq. \ref{eq:dir}). 
We found a maximum directivity of 300 related to an emission cone semi-angle of 7$^{\circ}$.  Following the principle of microwave HAs, the optical HA is able to efficiently convert dipolar emission into extremely directive radiations that can be efficiently outcoupled into free space mode continuum or discrete fiber modes.\\

\begin{figure}[htbp]
\begin{center}
\includegraphics [width=0.98\columnwidth]{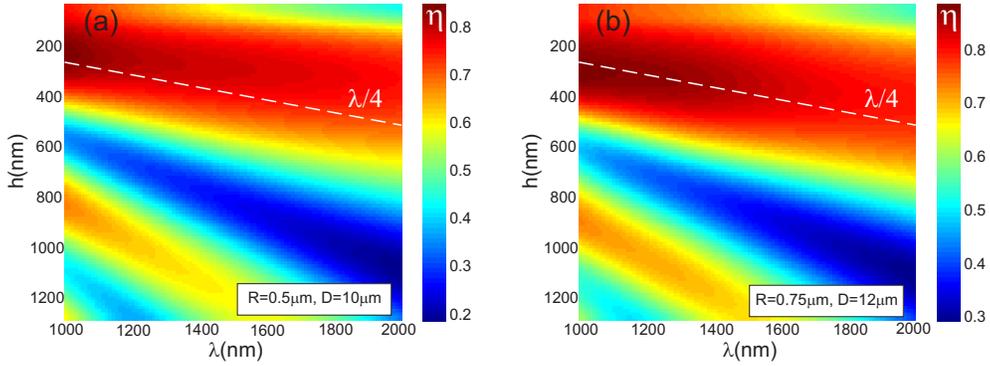}
\caption{(a,b) collection efficiency $\eta$ as a function of $\lambda$ and the dipole-to-mirror distance $h$, for the two different HA geometries considered in Fig. \ref{fig:colldir}(a) (detailed in the figure insets).}\label{fig:coll}
\end{center}
\end{figure}

Since the power collected by the HA is confined within the tapered waveguide (see Fig. \ref{fig:coll}(a)),  the collection efficiency $\eta$ (Eq. \ref{eq:eta}) will be defined as $P_{out}(r=D/2)/P_r$, ie. $P_i$ (Eq. \ref{eq:eta}) will be calculated by Poynting vector integration through the microtip output cross-section. Coefficient $\eta$ is reported in Figs. \ref{fig:coll}(a,b) as a function of the dipole-to-mirror spacing ($h$) for two HA geometries defined in the figure insets. The wavelength range spans in the near-infrared domain from 1000 nm to 2000 nm and the spacing $h$ ranges from 30 nm to 1270 nm.  The collection efficiencies of both HA geometries are oscillating functions of $h$ and their local maxima and minima are linear functions of $\lambda$. The spacing $h$ that induces maximum collection efficiency (called $h_{opt}$) is close to $\lambda/4$ for both HA geometries, which is consistent with the empirical rules of microwave HA design (the spacing $h=\lambda/4$ is represented in the figures with white dashed lines). The little mismatch between the optimum $h$ values achieved for the two HA configurations is the result of the complex optical mechanism between the dipolar source and the optical HA that is strongly dependent on the HA geometrical parameters. Note that the free space propagation phenomenon between the dipole and the mirror explains why the local maxima of $\eta$ are decreasing functions of $h$ and increasing functions of the ratio $R/\lambda$ and why the mismatch $|h_{opt}-\lambda/4|$ is a increasing function $R/\lambda$. It also explains why the optical HA has a spectrally broadband operation compared to the microwave HA for which waveguiding occurs  between the feed and the reflector. We see from Fig. \ref{fig:coll}(b) that the collection efficiency $\eta$ can stay above 78\% over the spectral domain 1000 nm-2000 nm with a maximum value larger than 88\% at $\lambda=1000$ nm. By comparing Figs. \ref{fig:colldir}(d) and \ref{fig:coll}, we see that the optical HA is capable of radiating in free space more than 85 \% of the energy from a dipole source within an emission cone semi-angle of 7$^{\circ}$ (corresponds to an antenna directivity of 300).

The outstanding collection and directivity performances of the optical HA could even be increased by optimizing the shape for the dielectric horn and/or using a parabolic mirror. They represent highly promising perspective in free space sensing applications and in the coupling of QEs to optical fibers. In the latter case, the directivity of the HA is not the only critical parameter, but the phase matching ability of the HA to optical fiber is also crucial.\\

\begin{figure}[htbp]
\begin{center}
\includegraphics [width=0.98\columnwidth]{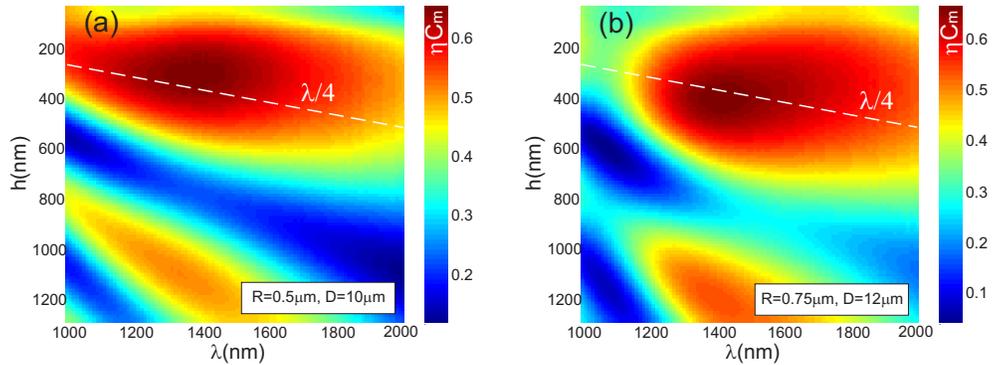}
\caption{(a,b) Part of the radiated power from the dipole source $\eta C_m$ that is collected and guided within the optical fiber by the two HA geometries detailed in the figure insets: coefficient $\eta C_m$ as a function of $\lambda$ and $h$.}\label{fig:final_dist}
\end{center}
\end{figure}

 The ability of the optical HA to outcouple the collected photons into a SMF-28 monomode optical fiber is presented in Fig. \ref{fig:final_dist}. This antenna property is described by coefficient $\eta C_m$ (see section 3), which represents the fraction of radiated power from the dipole source that is collected by the HA and guided within the optical fiber. Figs. \ref{fig:final_dist}(a) and (b) show $\eta C_m$ as a function of $\lambda$ and $h$, for the two different HA geometries shown previously (see figure insets). Our calculation does not take into account the free space propagating photons that are projected onto the entrance facet of the fiber and coupled to the fiber mode. Simulations that include this dipole-to-fiber coupling channel show discrepancies in the energy transfer which do not exceed 1\%. Therefore, $\eta C_m$ is a very good approximation of the probability that a photon radiated in free space by a QE coupled with the optical HA is guided into the fiber.

These results show that the maximum coupling efficiency between the dipole source and the optical fiber can exceed 63 \% (at $\lambda=1450$ nm) for the two particular HA geometries studied here. $\eta C_m$ remains slightly higher for the larger radius of curvature of the microtip (R=750 nm). Such energy transfers are made possible by the ability of the HA to direct the dipole emission and to project field distributions onto the fiber entrance facet that tightly overlap the fiber mode (gaussian-like profile of the proper size and plane wavefronts). For both HA geometries, the maximum dipole-to-fiber energy transfer is achieved for dipole-to-mirror spacings $h$ close to $\lambda/4$, which is consistent with the operating principle of HAs.\\

\begin{figure}[htbp]
\begin{center}
\includegraphics [width=0.98\columnwidth]{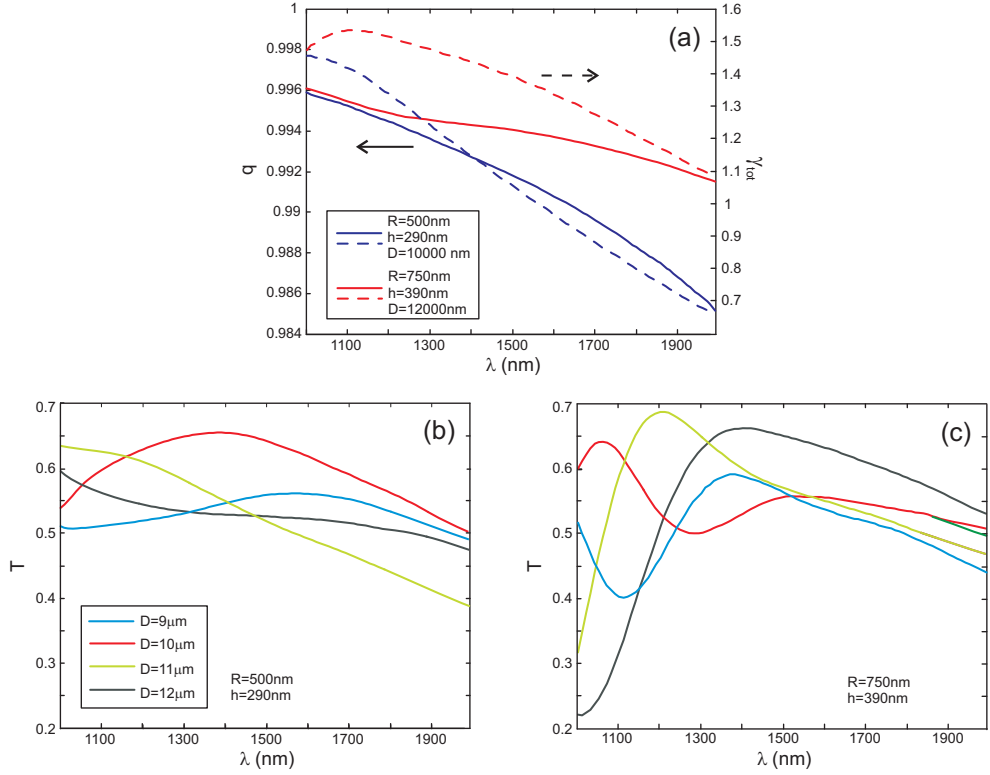}
\caption{(a) Spectra of total decay rate ($\gamma_{tot}$, dashed lines) and quantum yield ($q$, solid lines) of the structure for the two HA geometries detailed in the figure inset.(b) spectrum of the overall photon transfer $T$ from the QE to the optical fiber, for $R=500$nm, $h=290$nm and 4 different values of parameter $D$ ranging from 9000 nm to 12000 nm (see inset).(c) spectrum of the overall photon transfer  $T$ from the QE to the optical fiber, for $R=750$nm, $h=390$nm and 4 different values of parameter $D$ ranging from 9000 nm to 12000 nm (see inset of (b)).  }\label{fig:transmr}
\end{center}
\end{figure}

Figure \ref{fig:transmr}(a) reports the spectra of the total decay rate $\gamma_{tot}$ and the quantum yield $q$ of the HA for the two HAs considered here (see figure insets). Since the energy transfer does not rely on the interaction of the nano-emitter with a high quality factor and/or small mode volume  resonator, the emitter's decay rate is not strongly enhanced (dashed curves) and is rather a slowly increasing function of $h/\lambda$ (result consistent with Ref. \cite{vion:ox10}).  One main drawback of the optical HA is that it can induce lower emission rates than free space when $h/\lambda$ becomes too small. Fig. \ref{fig:transmr}(a) shows for example that $\gamma_{tot}<1$ when $\lambda > 1550$nm, for the HA geometry with $R=500$nm and $h=290$nm (blue dashed curve). Therefore, the radius of curvature at the microtip apex has to be chosen in a way that the optimum spacing $h_{opt}$ is as large as possible to avoid weak emission configurations. For example, the HA geometry with $R=750$nm does not hinder the emission rate of the QE over the spectral range considered in this study. The quantum yield (solid curves) of the structure is however almost unaffected (in opposition to metallic nano-antennas which dissipate optical energy) and the HA radiation process remains efficient over unprecedented spectral bandwidths of several hundreds nanometers. \\

Fig. \ref{fig:transmr} reports the overall photon transfer from the QE to the optical fiber as a function of the flaring angle of the dielectric waveguide.  Figs. \ref{fig:transmr}(b) and (c) show the spectra of parameter $T$, ie. the probability that a photon emitted by the QE is collected and guided within the optical fiber (see Eq. \ref{eq:T}), for two different pairs ($R,h$) given in the figure insets, respectively. For each pair ($R,h$), the diameter $D$ of the output interface of the flaring dielectric waveguide is varied from 9 $\mu$m to 12 $\mu$m by steps of 1 $\mu$m. $D$ has to be chosen to match the width of the field distribution that leaves the optical HA to the width of the fiber mode. We see that the influence of $D$, ie. the flare angle of the microtip, onto parameter $T$ is important. However, $D$ is not a predominant parameter in the HA-to-fiber coupling since the relationship between $D$ and $T$ is strongly dependant on the other geometrical $R$ and $h$. Therefore, a trade-off has to be found between the geometrical parameters $R$, $L$ and $D$ of the flaring dielectric waveguide to optimize the photon transmission in the optical fiber. When $R=$500nm, $h=$290nm and $D$=10 $\mu$m , $T$ remains above 50\% for wavelengths spanning between 1000 nm and 2000 nm and with a maximum value of about 65\% at $\lambda=$1400 nm (Fig. \ref{fig:transmr}(b)). When $R=$750nm, $h=$390nm and $D$=11 $\mu$m, $T$ is larger than 50\% for wavelengths ranging between 1000 nm and 2000 nm and reaches a maximum value of about 70\%  at $\lambda=$1220 nm. The HA configuration involving $R=$750nm and $D$=12 $\mu$m (gray curve of Fig. \ref{fig:transmr}(c)) is of particular interest for applications in telecommunications since it insures a ratio of guided photons into the fiber larger than 60\% over the spectral range 1280 nm-1750 nm with a maximum of $67\%$ at $\lambda=1430$nm.



\section{Conclusion}

In summary, the optical horn antenna combines high collection efficiency, unprecedented directivity and an excellent phase-matching ability with vacuum and optical fibers. Since our optical horn antenna is based on a lossless non-resonant optical mechanism, it can operate over remarkably broad spectral bandwidths while keeping a quantum yield above 0.99. Such properties do not seem to be achievable with plasmonic nano-antennas. Similar to the microwave HA, the optical HA needs to be large compared to the wavelength in order to provide its unique directivity and phase-matching properties. Therefore, it cannot compete, in terms of compactness, with resonant directive subwavelength nano-antennas as the Yagi-Uda \cite{li:prb07,hofmann:njp07,taminiau:ox08}. However, the development of fiber bright nano-emitters or fiber nanosensors relaxes the size requirements of the optical antenna which has to be attached onto the large scale fiber entrance facet. Therefore, the use of ultra compact resonant antennas is not necessary here. Furthermore, these structures do not seem to be well adapted to efficiently launch and phase-match photons to a fiber mode. In that context, the optical horn antenna may provide a promising solution for integrating quantum emitters to fiber networks for telecommunications, quantum optics and sensing applications. It can also achieve ultra directive emissions in free space which is attractive in the development of directed bright nanoscale light sources for quantum optics and sensing applications. For example, this work paves the way toward the generation of fiber-integrated on-demand single photon sources, for which the low-loss and wide-band photon transfer provided by the optical horn antenna are of particular interest.

\section*{Acknowledgements}

This work is funded by "Région Franche-Comté" under project AMIFI and by the "Labex ACTION". It is supported by the Pôle de Compétitivité Microtechnique.


\begin{thebibliography}{10}

\bibitem{miyazawa:jjap05}
T.~Miyazawa, K.~Takemoto, Y.~Sakuma, S.~Hirose, T.~Usuki, N.~Yokoyama,
  M.~Takatsu, and Y.~Arakawa,
\newblock "Single-photon generation in the 1.55-mum optical-fiber band from an
  inas/inp quantum dot,"
\newblock {Jpn J Appl Phys} \textbf{44}, L620--L622 (2005).

\bibitem{moreau:apl01}
E.~Moreau, I.~Robert, JM~G{\'e}rard, I.~Abram, L.~Manin, and V.~Thierry-Mieg,
\newblock "Single-mode solid-state single photon source based on isolated
  quantum dots in pillar microcavities,"
\newblock {Appl. Phys. Lett.} \textbf{79}, 2865--2867 (2001).

\bibitem{pelton:prl02}
M.~Pelton, C.~Santori, J.~Vuc?kovi{\'c}, B.~Zhang, G.S. Solomon, J.~Plant, and
  Y.~Yamamoto,
\newblock "Efficient source of single photons: a single quantum dot in a
  micropost microcavity,"
\newblock {Phys. Rev. Lett.} \textbf{89}, 233602 (2002).

\bibitem{claudon:natphot10}
J.~Claudon, J.~Bleuse, N.S. Malik, M.~Bazin, P.~Jaffrennou, N.~Gregersen,
  C.~Sauvan, P.~Lalanne, and J.M. G{\'e}rard,
\newblock "A highly efficient single-photon source based on a quantum dot in a
  photonic nanowire,"
\newblock {Nat. Photon.} \textbf{4}, 174--177 (2010).

\bibitem{gerard:josab09}
D.~G{\'e}rard, A.~Devilez, H.~Aouani, B.~Stout, N.~Bonod, J.~Wenger, E.~Popov,
  and H.~Rigneault,
\newblock "Efficient excitation and collection of single-molecule fluorescence
  close to a dielectric microsphere,"
\newblock {J. Opt. Soc. Am. B} \textbf{26}, 1473--1478 (2009).

\bibitem{devilez:nl10}
A.~Devilez, B.~Stout, and N.~Bonod,
\newblock "Compact metallo-dielectric optical antenna for ultra directional and
  enhanced radiative emission,"
\newblock {Nano Lett.} \textbf{4}, 3390--3396 (2010).

\bibitem{curto:science10}
A.G. Curto, G.~Volpe, T.H. Taminiau, M.P. Kreuzer, R.~Quidant, and N.F. van
  Hulst,
\newblock "Unidirectional emission of a quantum dot coupled to a nanoantenna,"
\newblock {Science} \textbf{329}, 930--933 (2010).

\bibitem{lee:natphot11}
KG~Lee, XW~Chen, H.~Eghlidi, P.~Kukura, R.~Lettow, A.~Renn, V.~Sandoghdar, and
  S.~G{\"o}tzinger,
\newblock "A planar dielectric antenna for directional single-photon emission
  and near-unity collection efficiency,"
\newblock {Nat. Photon.} \textbf{5}, 166--169 (2011).

\bibitem{chen:ol11}
X.W. Chen, S.~G{\"o}tzinger, and V.~Sandoghdar,
\newblock "99\% efficiency in collecting photons from a single emitter,"
\newblock {Opt. lett.} \textbf{36}, 3545--3547 (2011).

\bibitem{chang:prl06}
D.E. Chang, A.S. S\o{}rensen, P.R. Hemmer, and M.D. Lukin,
\newblock "Quantum optics with surface plasmons,"
\newblock {Phys. Rev. Lett.} \textbf{97}, 053002 (2006).

\bibitem{davanco:ol09}
M.~Davan{\c{c}}o and K.~Srinivasan,
\newblock "Fiber-coupled semiconductor waveguides as an efficient optical
  interface to a single quantum dipole,"
\newblock {Opt. lett.} \textbf{34}, 2542--2544 (2009).

\bibitem{balanis}
C.A. Balanis,
\newblock {\em Antenna theory: analysis and design}
\newblock (John Wiley \& Sons, New-York, 1997).

\bibitem{wade:06}
P.~Wade,
\newblock "Rectangular waveguide to coax transition design,"
\newblock {http://f1chf.free.fr/PDF/convertisseurs WR90 et WR75.pdf}.

\bibitem{lukosz:josa77}
W.~Lukosz and R.~Kunz,
\newblock "Light emission by magnetic and electric dipoles close to a plane
  dielectric interface. ii. radiation patterns of perpendicular oriented
  dipoles,"
\newblock {J. Opt. Soc. Am.} \textbf{67}, 1615--1619 (1977).

\bibitem{taflove:book}
A.~Taflove and S.C. Hagness,
\newblock {\em Computational Electrodynamics: The Finite-Difference Time-Domain
  Method, Third Edition}
\newblock (Artech House, Boston, 2005).

\bibitem{novotny:nanooptics}
L.~Novotny and B.~Hecht,
\newblock {\em Principle of nano-optics}
\newblock (Cambridge University Press, 2006).

\bibitem{palik:book}
E.D. Palik,
\newblock {\em Handbook of optical constants of solids}
\newblock (Academic Press, 1998).

\bibitem{thomson:83}
Thomson-CSF,
\newblock {\em L'optique guid\'ee monomode et ses applications}, \textbf{15}
\newblock (Masson, 1983).

\bibitem{vion:ox10}
C.~Vion, P.~Spinicelli, L.~Coolen, C.~Schwob, J.M. Frigerio, J.P. Hermier, and
  A.~Ma{\^\i}tre,
\newblock "Controlled modification of single colloidal cdse/zns nanocrystal
  fluorescence through interactions with a gold surface,"
\newblock {Opt. express} \textbf{18}, 7440--7455 (2010).

\bibitem{li:prb07}
J.~Li, A.~Salandrino, and N.~Engheta,
\newblock "Shaping light beams in the nanometer scale: A yagi-uda nanoantenna in
  the optical domain,"
\newblock {Phys. Rev. B} \textbf{76}, 245403 (2007).

\bibitem{hofmann:njp07}
H.F. Hofmann, T.~Kosako, and Y.~Kadoya,
\newblock "Design parameters for a nano-optical yagi--uda antenna,"
\newblock {New J. Phys.} \textbf{9}, 217 (2007).

\bibitem{taminiau:ox08}
T.H. Taminiau, F.D. Stefani, and N.F. van Hulst,
\newblock "Enhanced directional excitation and emission of single emitters by a
  nano-optical yagi-uda antenna,"
\newblock {Opt. Express} \textbf{16}, 10858--10866 (2008).

\end{thebibliography}
\end{document}